\documentclass[fleqn,usenatbib]{mnras}

\usepackage[brazilian]{babel}
\usepackage[utf8]{inputenc}

\usepackage{graphicx}	
\usepackage{amsmath}	
\usepackage{amssymb}	

\newcommand{\feh}{[Fe/H]}
\newcommand{\afe}{[$\alpha$/Fe]}
\newcommand{\adev}{$\Delta_{\rm \lambda}$}

\newcommand*{\orcid}{\includegraphics[scale = 0.0525]{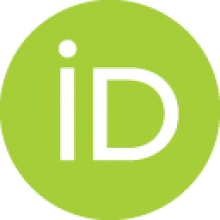}}

\title[How well spectral fitting performs?]{How well can we determine ages and chemical abundances from spectral fitting of integrated light spectra?}

\author[Gonçalves et al.]
{\href{https://orcid.org/0000-0003-4999-3691}{Geraldo Gonçalves\orcid}$^{1}$\thanks{E-mail: geraldo.goncalves.santos@usp.br}, \href{https://orcid.org/0000-0003-1846-4826}{Paula Coelho\orcid}$^{1}$, \href{https://orcid.org/0000-0002-2244-0897}{Ricardo Schiavon\orcid}$^{2}$, \href{https://orcid.org/0000-0002-7383-7106}{Christopher Usher\orcid}$^{2}$\\
$^{1}$Instituto de Astronomia, Geofísica e Ciências Atmosféricas, Universidade de São Paulo, São Paulo, SP, Brazil\\
$^{2}$Astrophysics Research Institute, Liverpool John Moores University, 146 Brownlow Hill, Liverpool L3 5RF, UK
}

\date{Accepted XXX. Received YYY; in original form ZZZ}

\pubyear{2019}

 \hypersetup{draft}
\begin{document}
\label{firstpage}
\pagerange{\pageref{firstpage}--\pageref{lastpage}}
\maketitle

\begin{abstract}
The pixel-to-pixel spectral fitting technique is often used in studies of stellar populations. It enables the user to infer several parameters from integrated light spectra such as ages and chemical abundances. In this paper, we examine the question of how the inferred parameters change with the choice of wavelength range used.
We have employed two different libraries of integrated light spectra of globular clusters (GCs) from the literature and fitted them to stellar population models using the code  \textsc{Starlight}. We performed tests using different regions of the spectra to infer reddening, ages, [Fe/H] and [$\alpha$/Fe]. Comparing our results to age values obtained from isochrone fitting and chemical abundances from high resolution spectroscopy, we find that: (1) The inferred parameters change with the wavelength range used; (2) The method in general retrieves good reddening estimates, specially when a wider wavelength range is fitted; (3) The ideal spectral regions for determination of age, [Fe/H], and [$\alpha$/Fe] are: 4170-5540\AA, 5280-7020\AA, and 4828-5364\AA, respectively; (4) The retrieved age values for old metal-poor objects can be several Gyr younger than those resulting from isochrone fitting. We conclude that, depending on the parameter of interest and the accuracy requirements, fitting the largest possible wavelength range may not necessarily be the best strategy.
\end{abstract}




\section{Introduction}
Determining realistic values for ages and chemical abundances of stars is a very important step to understand how stellar systems such as galaxies and stellar clusters form and evolve. For the Milky Way, as we are able to observe stars individually, we can use isochrone fitting on the Color-Magnitude Diagram (CMD) of stellar clusters \citep[e.g.][]{2010ApJ...708..698D} or calibrations of asteroseismologic masses \citep[e.g.][]{ness+16} to determine ages, and high resolution spectroscopy to derive abundances \citep[e.g.][]{barbuy+18}. But for more distant systems, resolving stars become increasingly more difficult and we have to work with integrated light.

There is a range of techniques that can be used to infer these parameters from integrated light, from the widely used historical line indices (e.g. \citealt{worthey+94idx, trager+00, thomas+03, schiavon+07}) to updated approaches such as SED fitting \citep[e.g.][]{sed}, Principal Component Analysis \citep[e.g.][]{pca}, Bayesian inference \citep[e.g.][]{mcmc}, pixel-to-pixel spectral fitting (e.g. \citealt{cid07, koleva+08, walcher+09, conroy+14}), among others.

In this paper, we are concerned with spectral fitting. One aspect which has not been studied in detail is to quantify to what extend the choice of the wavelength range affects the results of the spectral fitting. The literature reports conflicting results. 

\citet{koleva+07} estimated the star formation history of galaxies using models computed with PEGASE.HR \citep{leborgne+04} and a parametric procedure.
They concluded that the wavelength range used in the fit does not affect significantly the precision of derived kinemactics parameters and metallicities. However, the precision of ages determinations became worse when smaller wavelength ranges were used.

In \citet{koleva+08}, the authors used two different algorithms -- \textsc{STECKMAP} \citep{ocvirk+06} and \textsc{NBURSTS} \citep{chilingarian+07} -- and three libraries of Simple Stellar Populations (hereafter SSP) -- ELODIE 3.1 \citep{prugniel+07}, Galexev (\citealt{bc+03}, commonly called BC03) and MILES (\citealt{vazdekis+99}, \citealt{vazdekis+03}) -- to fit GCs from \citet{schiavon+05} and the integrated spectrum of M67 from \citet{schiavon+04}. They performed the fit in the 4000-5700 \AA\ region, successfully retrieving the age and metallicities parameters for 36 out of 41 clusters. But they briefly reported avoiding the bluest region of the spectra in the fit (in particular the H\&K lines), otherwise the inferred ages would be systematically lower by $\sim$\,1\,Gyr.

\citet{walcher+09} used the \textsc{sedfit} algorithm \citep{walcher+06} to fit spectra of NGC 6528 and NGC 6553 from \citet{schiavon+05}. They tested five different wavelength ranges and found that 4828--5364 \AA\ is the one that results in [Fe/H], [$\alpha$/Fe] and age values closer to the reference values from literature. The other four ranges seem more affected by the age-[Fe/H] degeneracy. 

\citet{cezario+13} used two different samples of globular clusters: M31 GCs from \citet{alves-brito+09} and Galactic GCs from \citet{schiavon+05}. Using the \textsc{ULySS} algorithm \citep{koleva+09} and SSP models from \citet{vazdekis+10}, they tested four different wavelength ranges with the Galactic GCs. They concluded that this choice has little influence on the metallicities, while the ages can vary a lot: a third of the spectra were fitted with intermediate ages in some cases. They favour 4000-5400 \AA, being the one that best reproduced CMD ages, and applied the technique to the GCs in M31.

In this work, we examine this question by performing assesing the dependence of the retrieved stellar population parameter with the wavelength range used.  We aim at finding what would be the ideal configuration (if any) to infer physically consistent parameters, for a fixed choice of spectral fitting code and SSP models. Thus, we hope this work can contribute to studies of integrated light, its properties and methods in order to infer accurate parameters.  

This paper is organized as follows: Section \ref{obsdata} describes the data from literature we used, section \ref{methodology} describes the methodology, section \ref{results_chapter} presents our results, that are discussed in detail in section \ref{sec_discussions}. Finally, section \ref{sec_conclusion} presents our conclusions.

\section{The Observational Data}
\label{obsdata}

\begin{figure}
	\includegraphics[width=\columnwidth]{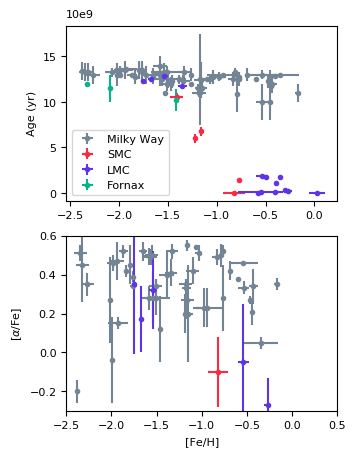}
    \caption{Distributions of age and abundances for the globular clusters in WAGGS data. The upper panel shows age versus [Fe/H], and lower panel shows [$\alpha$/Fe] versus [Fe/H]. Host galaxies are distinguished by colour, as indicated in the panel.}
    \label{fig_waggs}
\end{figure}

In this study, we have used data from the WiFes Atlas of Galactic Globular clusters Spectra (WAGGS) project \citep{waggs1}. The WAGGS library contains objects from the Milky Way, the Magellanic Clouds and the Fornax galaxy, thus covering a wide range in ages and chemical abundances (see Figure \ref{fig_waggs}).

The data was taken with the WiFeS integral field spectrograph \citep{dopita+07,dopita+10} on the Australian National University 2.3m telescope.
The spectra were taken in four different arms, using different grating setups: 
U7000, covering 3270--4350 \AA\ range;
B7000, covering 4170--5540\,\AA;
R7000 covering 5280--7020\,\AA, and;
I7000 covering 6800--9050\,\AA. 
All four gratings give spectral resolutions of $\delta \lambda / \lambda \sim 6800$. The data is publicly available at the WAGGS project website\footnote{\url{http://www.astro.ljmu.ac.uk/~astcushe/waggs/}}.

\begin{figure*}
\centering
	\includegraphics[scale=0.9]{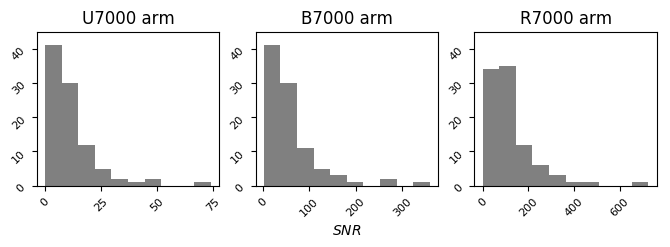}
    \caption{Distribution of signal-to-noise ratios (SNRs) of the spectra in the WAGGS library. The three panels show histograms of SNRs for the U7000, B7000 and R7000, respectively. It is noticeable that U7000 show the lowest values of SNR of the sample.}
    \label{snr}
\end{figure*}

The error spectra are provided, as computed from the pipeline \texttt{PYWIFES} \citep{childress+14}. The median SNR for each arm are 12, 61 and 124 for the U, B and R arms respectively\footnote{In this work we have not used the I7000 arm because the SSP models employed do not cover its spectral region}. The distribution of the SNR values per arm are shown in Figure \ref{snr}. Given the natural difficulties of calibrating IFU data, these errors are likely overestimated at low SNR and underestimated at high SNR, and therefore may be biased. In particular, at high SNR, the dominant source of uncertainty is effects stochastic due to the finite stellar mass within the observed field of view.

For some additional analysis, we also used GC integrated spectra from \citet{schiavon+05}. These data have a different coverage range (3350--6430\AA), with a resolution of 3.1\,\AA\ (FWHM). 
Each observation was taken by drifting the spectrograph slit across the core diameter of the object in exposures of 15 minutes. The data are publicly available at the National Optical Astronomical Observatory website\footnote{\url{ http://www.noao.edu/ggclib}}.



In order to obtain the reference values for ages and chemical abundances for the globular clusters, we have used the Table 1 in \citet{waggs1} as a starting point, and performed an extensive search in literature. 
Searching object by object, we gathered information from literature prioritising works that provide [$\alpha$/Fe] (here represented by [Mg/Fe]) as well as [Fe/H] and their uncertainties. This compilation is shown in Table \ref{ref_table}.

\begin{table*}
\caption{\label{ref_table}[Abridged; full table is available as online-only material] Globular clusters present in the WAGGS database and the reference values for ages and chemical abundances. The symbol '*' indicates values that were not found in our search.}
\begin{tabular}{llllcclcl}
\hline
ID      & Galaxy & RA {[}$^o${]} & Dec {[}$^o${]} & {[}Fe/H{]} & {[}Mg/Fe{]} & Reference for abundances                                 & Age (Gyrs)   & Reference for age    \\
\hline
NGC0104 & MW     & 6.024      & -72.081     & -0.768  $\pm$    0.031         & 0.52    $\pm$     0.03           & \citet{2009AeA...505..139C}  & 12.75 $\pm$ 0.5    & \citet{2010ApJ...708..698D}                        \\
NGC0121 & SMC    & 6.701      & -71.536     & -1.41   $\pm$    0.07          & *       $\pm$     *              & \citet{2004oee..sympE..29J}  & 10.5  $\pm$ 0.5    & \citet{2008AJ....135.1106G}                        \\
NGC0330 & SMC    & 14.074     & -72.463     & -0.82   $\pm$    0.11          & -0.1    $\pm$     0.18           & \citet{1999AeA...345..430H}  & 0.03  $\pm$ 0.002  & \citet{2002ApJ...579..275S}                        \\
NGC0362 & MW     & 15.809     & -70.849     & -1.17   $\pm$    0.05          & 0.33    $\pm$     0.04           & \citet{2013yCat..35570138C}  & 11.5  $\pm$ 0.5    & \citet{2010ApJ...708..698D}      \\
\hline
\end{tabular}
\end{table*}





\section{Methodology}\label{methodology}

\subsection{\label{sec_prepross}Pre-processing of the data}

Prior to the stellar population analysis, the data had to be processed to be in the rest-frame and the spectral resolution adjusted.

We measured the radial velocities of the observations using the task \texttt{fxcor} in \texttt{IRAF} \footnote{\url{http://ast.noao.edu/data/software}}. Each observation was cross-correlated to 3 templates, representing stars with different spectral types, obtained from the theoretical spectral library from \citet{coelho14}. The atmospheric parameters (T$_{\rm eff}$, $\log g$, \feh, \afe) of the adopted templates were: (4250, +2.5, -1.0, +0.4); (5750, +4.0, --0.5, +0.0), and (10000, +2.0, --0.5, +0.0). We adopted as final radial velocity in each observation the cross-correlation with the template which resulted in the smallest estimated error. The radial velocities from the 3 different templates were similar for most of the sample, and the adopted values are available in the online material. We verified that our measured values were in good agreement with heliocentric radial velocities from \citealt{globclust} (2010 edition). For the U7000 arm, we could not measure the radial velocities for 9 clusters due to the low SNR. 

After correcting the spectra to the rest-frame, we altered their spectral resolution to match the one of the models we used (Section \ref{sec_specfit}).
The resolution as function of radius was measured reducing arc exposures in the same manner as science exposures, then fitting arclines in the resulting 1d spectra with Gaussians. The measurements of individual arc lines were then fit with quadratic polynomials to give the following relations.

\begin{equation}
    R_{U7000} \approx -3.14 \times 10^{-3} \lambda^2 + 2.70\times 10^1 \lambda - 5.09
    \label{resolution_U}
\end{equation}{}
\begin{equation}
    R_{B7000} \approx 7.50\times 10^{-4} \lambda^2 - 5.93 \lambda + 1.74\times 10^4
    \label{resolution_B}
\end{equation}{}
\begin{equation}
    R_{R7000} \approx 1.39\times 10^{-4} \lambda^2 - 1.94\times 10^{-1} \lambda + 2.75\times 10^3
    \label{resolution_R}
\end{equation}{}
\begin{equation}
    R_{I7000} \approx -2.04\times 10^{-7} \lambda^2 + 8.14\times 10^{-1} \lambda + 4.10\times 10^2
    \label{resolution_I}
\end{equation}{}

We applied a gaussian pixel-to-pixel convolution on the spectra using the functions \texttt{gaussian} and \texttt{convolve} from \textsc{Python}'s \texttt{scipy} package. We used the equations \ref{resolution_U} to \ref{resolution_R} to match the data to a resolution of FWHM = 2.51\,\AA\ \citep[][]{fbarroso+11}\footnote{We have not used spectra observed in the I7000 arm in the present work, but report the obtained relations for completeness.}.

In order to fit the spectra in the $R_M$ range, we concatenated the spectra, prioritizing the arm with greater SNR (i.e. using the data with greater SNR in the regions where the arms overlap). We also excluded the 8 first and the 8 last pixels of each arm to avoid edge effects. 

\subsection{\label{sec_specfit}Spectral fitting}

In this work, we have used the software \textsc{Starlight}  \citep{cid+05}, a full spectral fitting code widely used in galaxy studies. It creates a model $M$ fitting an observed spectrum as a combination of SSP models, as shown in Equation \ref{model_starlight}.

\begin{center}
\begin{equation}
M = \sum_{j}^{N_{\star}} x_j \mbox{SSP}_{j} \otimes G(v_{\star},\sigma_{\star}) \ 10^{-0.4 A_{V}}
\label{model_starlight}
\end{equation}
\end{center}

\noindent where $x_j$ is the population vector (it corresponds to the light contribution in percentage of each $\mbox{SSP}_j$ in the model), $G$ is the kinematical filter (where $v_{\star}$ is the radial velocity and $\sigma_{\star}$ is the velocity dispersion) and $A_V$ is the reddening (in magnitudes in the $V$ band).

Fitting this model allows us to infer light-weighted and mass-weighted ages and chemical abundances, as shown in Equations \ref{age_log}, \ref{feh} and \ref{alpha}, alongside reddening values and the kinematical parameters. In the present work, we fixed the kinematics as we already have the spectra in the rest frame and with the broadening matching that of the SSP models (see Section \ref{sec_prepross}) Although the clusters have their own velocity dispersion (about 5 to 15 km/s, \citealt{dalgleish+20}), these are significant smaller than the resolution of the MILES library.

\begin{center}
\begin{equation}
\langle\log({age})\rangle = \sum_{j}{x_j \cdot \log({age})_j}
\label{age_log}
\end{equation}
\end{center}

\begin{center}
\begin{equation}
\langle{\rm [Fe/H]}\rangle = \sum_{j}{x_j \cdot {\rm [Fe/H]}_j}
\label{feh}
\end{equation}
\end{center}

\begin{center}
\begin{equation}
\langle{\rm [\alpha/Fe]}\rangle = \sum_{j}{x_j \cdot {\rm [\alpha/Fe]}_j}
\label{alpha}
\end{equation}
\end{center}

When it comes to mass weighted parameters $x_j$ is replaced by $M\_cor_j$, which is the mass-weight of the $\mbox{SSP}_j$ in the model, corrected for mass lost during its evolution.

\textsc{Starlight} also returns the $\chi^2$ of each SSP fitted on the data, so we are able to work both with multi-population and simple-population fits.

The SSPs come from the MILES library \citep{vazdekis+15}, which uses the BaSTI isochrones (\citealt{basti+04, basti+06}). Each model is defined by three values: age (varying from 0.03 to 14.00 Gyrs), \feh\ (-2.27 to 0.26 dex) and \afe\ (0.0 and 0.4 dex). 
The models with \afe\,=0.0 adopt the solar abundances from \citet{grevesse+98}, scaled according to \feh. The $\alpha$-enhanced models adopt abundances of [X/Fe] = 0.4 for O, Ne, Mg, Si, S, Ca and Ti, and solar values for the other elements.

The library contains more than one thousand models, but we have selected for use 324 models, those with ages: 0.03, 0.06, 0.09, 0.20, 0.35, 0.50, 0.80, 1.25, 2.00, 2.75, 3.50, 4.50, 6.00, 7.50, 9.00, 10.50, 12.00 and 13.50 Gyr (one out of every three ages in the complete library). The spectral range goes from 3540.5 to 7409.6 \AA\ at a resolution of FWHM = 2.51\,\AA\ \citep{fbarroso+11}.

Given that the lowest age in the library is of 30\,Myr, we did not include the GCs NGC2004 and NGC2100 in the analysis, as they are reported in literature to be younger than this limit (Table \ref{ref_table}).

\subsection{A note on the definition of ages}\label{different_ages}


As reference values, we adopt ages derived from isochrone fitting to high quality CMD, compiled from literature (see Table \ref{ref_table}). But when inferring ages from integrated spectra, several definitions of an \emph{age} can be used. Below we shortly specify the different definitions used in this work.


\paragraph*{Isochrone age $t_{\rm iso}$:}  The age inferred from isochrone fitting, obtained from fitting stellar evolution isochrones to observed CMD. This is the value we adopt throughout this work as reference value for age. 

\paragraph*{Light-weighted age $t_{\rm light}$:} A multi-population mean age, computed by weighting the age of each SSP$_j$ by its light contribution $x_j$, as obtained from \textsc{Starlight}. This is the definition shown in Equation \ref{age_log}. 


\paragraph*{Mass-weighted age $t_{\rm mass}$:} A multi-population mean age, computed by weighting the age of each SSP$_j$ by its stellar mass. The mass fraction in stars for each SSP needs to be provided to \textsc{Starlight} to transform the light contribution $x_j$ into mass contribution $M_{{\rm cor}_j}$.



\paragraph*{SSP-equivalent age $t_{\rm SSP}$ :} Age of the SSP that best fits the observation (SSP model with the smallest $\chi^{2})$.

\paragraph*{Age of the top contributor to the integrated light $t_{{\rm max}\\, x_j}$:} The age of the SSP which most contribute to the integrated light (max(x$_j$)).\\

Equivalent definitions are valid for \feh\, and \afe, but our tests show that variations among the definitions are larger in the case of ages,  as discussed in Section \ref{results_chapter}.

\section{Results} \label{results_chapter}

For each cluster, we run \textsc{Starlight} in five different wavelength ranges, as listed in Table \ref{tab_ranges}. In Table \ref{features}, we list some of the main spectral features in each range. They include the four strongest lines of the Balmer series (H$\alpha$, H$\beta$, H$\gamma$, H$\delta$), the Balmer break, and strong the optical absorption features \citep{worthey+94idx, worthey_ottaviani97,1997A&A...325.1025P}.

In this section we present the results of the spectral fits, and their implications are discussed further in Section \ref{sec_discussions}.


\begin{table*}
\caption{Wavelength intervals used in this work for the spectral fitting of the globular clusters spectra.}
\begin{tabular}{lllll}
    \hline
     \textbf{Label} & \textbf{Interval (\AA)} & \textbf{Notes} & \multicolumn{2}{c}{\textbf{Normalization wavelengths (\AA)}}\\
      & & & For base spectra & For observed spectrum  \\ 
     \hline
     $R_U$  &  3540.5 -- 4350.0 & Intersection between U7000 and MILES range. & 4250.0 & 4240.0 -- 4260.0 \\ 
     $R_B$  &  4170.0 -- 5540.0 & Same as B7000 range. & 4250.0 & 4240.0 -- 4260.0  \\
     $R_R$  &  5280.0 -- 7020.0 & Same as R7000 range. & 7000.0 & 6990.0 -- 7010.0\\
     $R_W$ &  4828.0 -- 5364.0 & Interval favoured by \citet{walcher+09}. & 5050.0 & 5040.0 -- 5060.0 \\
     $R_M$ &  3540.5 -- 7409.6 & Same as MILES range. & 4250.0 & 4240.0 -- 4260.0 \\
     \hline
     \label{tab_ranges}
\end{tabular}
\end{table*}

\begin{table*}
\caption{Main spectral features in each wavelength range tested in this work.}
\begin{tabular}{lll}
    \hline
     \textbf{Label} & \textbf{Interval (\AA)} & \textbf{Main spectral features}\\
     \hline
     $R_U$  &  3540.5 -- 4350.0 & H$\gamma$, H$\delta$, CN$_1$, CN$_2$, Balmer break \\ 
     $R_B$  &  4170.0 -- 5540.0 & H$\beta$, H$\gamma$, Mgb triplet, Fe4383, Ca4455, Fe4531, Fe4668, Fe5015, Fe5270, Fe5335, Fe5406 \\
     $R_R$  &  5280.0 -- 7020.0 & H$\alpha$, TiO$_1$, TiO$_2$, Fe5709, Fe5782, Na D\\
     $R_W$ &  4828.0 -- 5364.0 & Mgb triplet, Fe5015, Fe5270, Fe5335 \\
     $R_M$ &  3540.5 -- 7409.6 & all listed above \\
     \hline
     \label{features}
\end{tabular}
\end{table*}

\subsection{Quality of the spectral fits}

The quality of the spectral fitting as performed by \textsc{Starlight} is given in terms of $\chi^{2}$ and the mean relative difference between model and observation \adev:

\begin{equation}
\Delta_{\rm \lambda} = \frac{1}{N} \sum_{\lambda}{\left| \frac{f_{\rm model}(\lambda)-f_{\rm obs}(\lambda)}{f_{\rm obs(\lambda)}}\right|}
\end{equation}

\noindent where \textit{N} is the number of pixels, $f_{\rm model}$ is the fitted spectrum and $f_{\rm obs}$ is the observed spectrum. 

\begin{figure*}
\centering
	\includegraphics[scale=0.9]{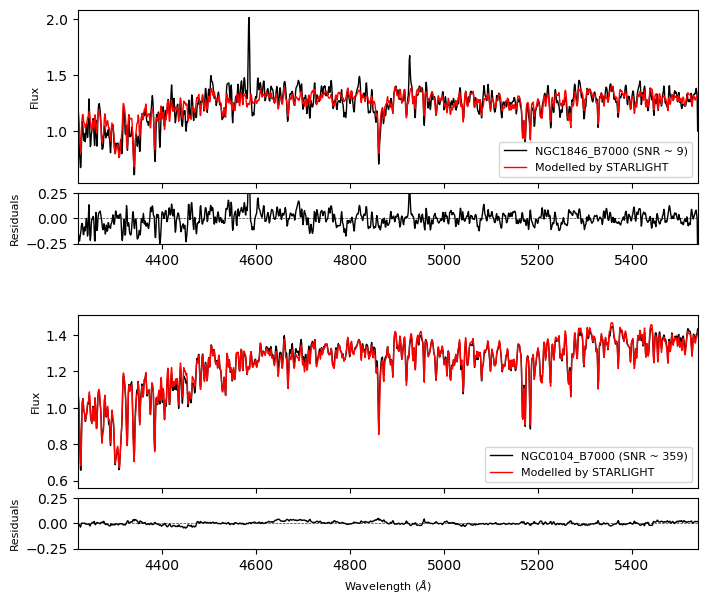}
    \caption{Examples of spectral fits performed with \textsc{Starlight}. The observed spectra from the WAGGS library are shown in black and the models fitted by \textsc{Starlight} in red. The residuals (observed flux - synthetic flux versus wavelength) are shown under each panel. The GCs displayed are NGC1846 (upper panel) and NGC0104 (lower panel).}
    \label{fit_NGC0104}
\end{figure*}

A good fit is illustrated in Figure \ref{fit_NGC0104}, for the case of NGC104 (47 Tuc), as observed in the B7000 arm in the WAGGS library ($\chi^2 \sim 2.6$, $\Delta_{\lambda} \sim 0.4$). 

\begin{figure*}
\centering
	\includegraphics[scale=0.75]{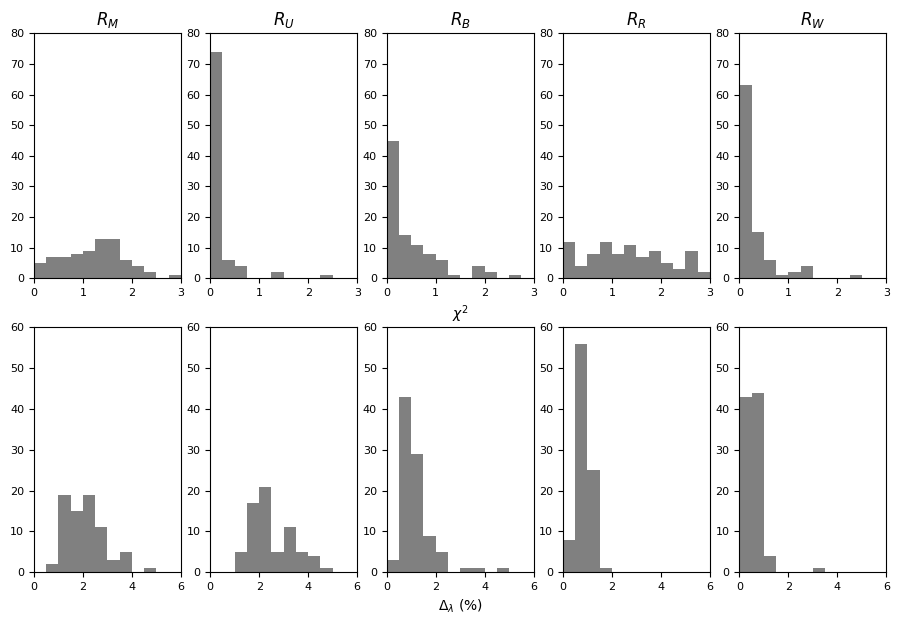}
    \caption{Distributions of $\chi^2$ (upper line) and $\Delta_{\lambda}$ (lower line) for the spectra of the WAGGS library. The columns show the histograms for the five different ranges fitted (see Table \ref{tab_ranges}). In the case of $\Delta_{\lambda}$, we show only the values up to 5. For the following cases, $\Delta_{\lambda}$ are not shown in the figure:  NGC1846-$R_M$ ($\Delta_{\lambda} \sim 28.3$), NGC6440-$R_M$ ($\Delta_{\lambda} \sim 31.5$), NGC6553-$R_M$ ($\Delta_{\lambda} \sim 24.1$), NGC6528-$R_M$ ($\Delta_{\lambda} \sim 31.9$), NGC2009-$R_U$ ($\Delta_{\lambda} \sim 21$), NGC6553-$R_U$ ($\Delta_{\lambda} \sim 51$), NGC6440-$R_U$ ($\Delta_{\lambda} \sim 55$), NGC1846-$R_U$ ($\Delta_{\lambda} \sim 98$) and NGC6528-$R_U$ ($\Delta_{\lambda} \sim 115$).}
    \label{plot_snr_adev}
\end{figure*}

In Figure \ref{plot_snr_adev} we illustrate the histograms of $\chi^2$ and \adev\ for all the fits performed. We only show \adev\ values up to 5, but for a few fits, they can go as high  as $\sim$18\% for R7000 and  $\sim$500\% for U7000 (both cases for NGC2004 spectra, which are very noisy with SN: < 1.0 in both cases
). All the cases where \adev\ > 5 are listed in the caption of Figure \ref{plot_snr_adev}. Overall, most spectra in the U arm are fitted within 2\% of the flux, and most spectra of B and R-arms are fitted within 1\%.


Another aspect of the fit quality has been discussed in \citet{cid+18}. The article discusses a way to identify if the algorithm performed the fit well or if the Markov-Chain did not converge. To verify this, the user can compare the $\chi^2_{{\rm SSP}_j}$ (the $\chi^2$ of fitting each SSP in the library) with the global $\chi^2$ of the multi-population fit. According to the author, in the case of a good fit  $\chi^2$ should be lower than any of the $\chi^2_{{\rm SSP}_j}$ values. Otherwise, the fit should not be trusted. Looking at these values, we found that only three of our fits (Fornax3--U7000, Fornax5--R7000 and NGC7099--B7000) do not meet this quality requirement. 


The cases which fall into the criteria above ($\chi^2 > \min(\chi^2_{\rm SSP})$) and/or were fitted with $\Delta_\lambda > 5$ have been removed from the analysis hereafter.

\subsection{Reddening}\label{ages_av}
Ages and reddening are the two parameters which primarily affect the continuum shape of the fitting process. As such, they are the ones likely to be more affected by uncertainties in flux calibration.

\begin{figure*}
\centering
	\includegraphics[scale=0.8]{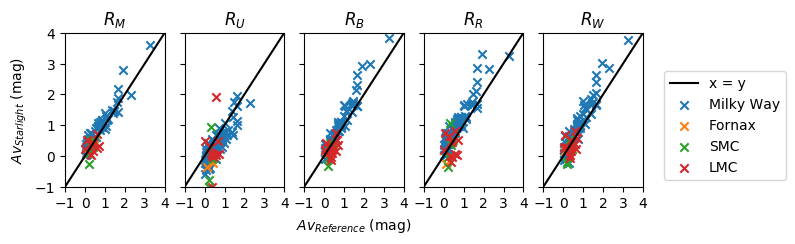}
    \caption{$A_V$ values from \textsc{Starlight} versus reference values. Each panel shows the results for one of the wavelength ranges tested in this work (see Table \ref{tab_ranges}). The identity line x=y is shown in solid black to guide the eye. Markers are colored by host galaxy, as indicated in the panel in the right.}
    \label{av_waggs_ref}
\end{figure*}

Figure \ref{av_waggs_ref} shows the redenning values $A_V$ inferred by \textsc{Starlight} compared to the reference values from literature (\citealt{globclust, mclaughlin+05}). For the majority of the fits, the correspondence is close to the 1-to-1 line. This good agreement between the fitted values and the literature values would seem to indicate that the relative WAGGS flux calibration is reliable. The average difference between \textsc{Starlight} and reference values are 0.10 for $R_M$, -0.25 for $R_U$, 0.12 for $R_B$, 0.31 for $R_R$ and 0.14 for $R_W$. For spectra in the B and R arms, the retrieved $A_V$ of systems with $A_V \gtrsim 2$ show a tendency to be overestimated. 

In the fitting runs we allow negative values for reddening down to -1. \citet{galazi+05} and \citet{mateus+06} discuss some reasons (mainly related to inconsistencies in the base of SSP models) why this should be allowed (clearly the results would have no physical meaning in these cases). Most of these cases with \textsc{Av} < 0 appear in the U7000 arm panel, which are the spectra with lower SNR (see Figure \ref{snr}).


\subsection{Ages}

\begin{figure*}
\centering
	\includegraphics[scale=0.9]{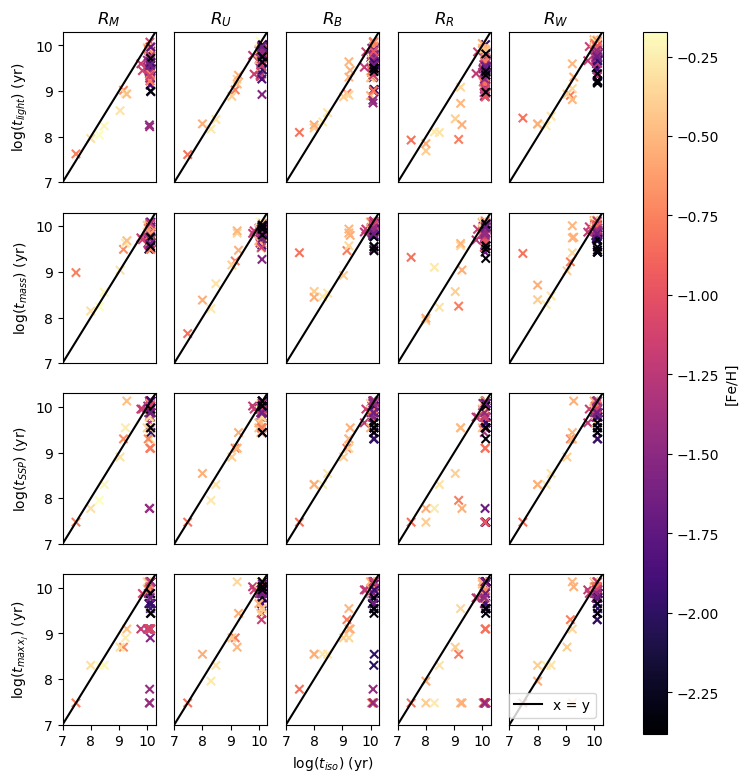}
    \caption{Ages retrieved from the spectral fitting (y-axis) versus isochrone ages (x-axis), in logarithmic scale. Each row shows the results for one of the different definitions of age (described in Section \ref{different_ages}). Each column represents one of the wavelength ranges tested in this work (see Table \ref{tab_ranges}). The identity line x=y is shown in solid black to guide the eye. Markers are colored by [Fe/H].}
    \label{ages_waggs_ref}
\end{figure*}

Figure \ref{ages_waggs_ref} shows the age values inferred from \textsc{Starlight} compared to the isochrone ages, for all definitions listed in Section \ref{different_ages}. For most of the young and intermediate-age objects, ages are generally well retrieved in spectra from the U7000 and B7000 arms, regardless of the age definition adopted. Values obtained from the R7000 arm spectra show larger dispersion, which is expected given that age features are clustered in the blue bands -- in the wavelength interval $R_R$ (see Table \ref{tab_ranges}) basically only H$_\alpha$ is available as an age feature.

\begin{figure*}
\centering
	\includegraphics[scale=0.9]{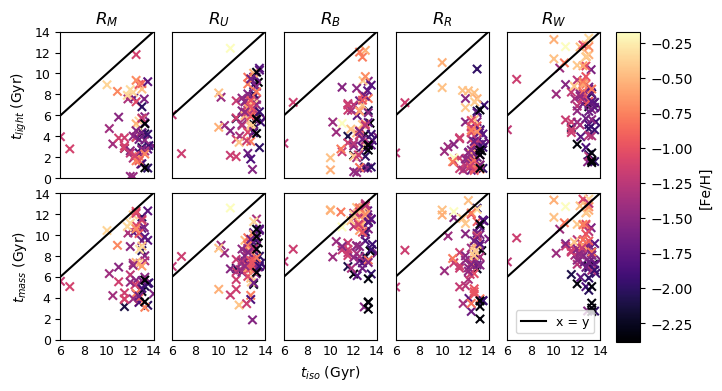}
    \caption{Retrieved values of age versus isochrone ages, in linear scale. Upper line shows the results of light-weighted ages, while the lower line shows the results of mass-weighted ages (see Section \ref{different_ages}). Each column represents one of the wavelength ranges tested in this work (see Table \ref{tab_ranges}). The identity line x=y is shown in solid black to guide the eye. Markers are colored by [Fe/H].}
    \label{ages_linear}
\end{figure*}

In the case of old systems, a range of ages is retrieved, with a tendency for the metal-poor objects to be modelled with ages younger than the isochrones ages. 
This is better illustrated in Figure \ref{ages_linear} showing $t_{\rm light}$ and $t_{\rm mass}$ in detail for the old systems. The difference between the spectroscopic age and the reference age can reach up to 10\,Gyr.  We further discuss these findings in Section \ref{sec_discussions}.

\subsection{Chemical abundances}\label{chemical_abundances}

\begin{figure*}
\centering
	\includegraphics[scale=0.9]{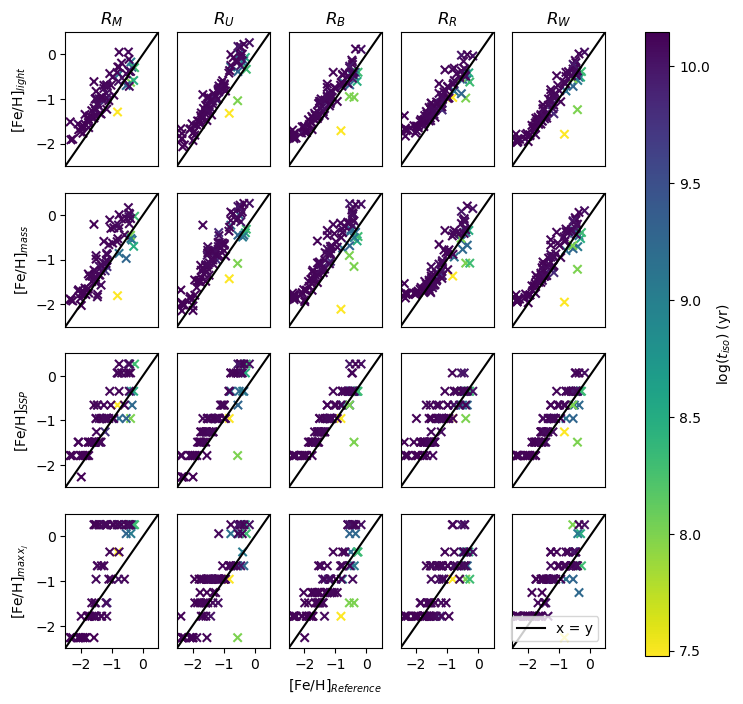}
    \caption{Retrieved values of [Fe/H] versus reference values. Each line shows the results for one of the different definitions of [Fe/H] (described in Section \ref{different_ages}). Each column represents one of the wavelength ranges tested in this work (see Table \ref{tab_ranges}). The identity line x=y is shown in solid black to guide the eye. Markers are colored by isochrone age.}
    \label{feh_waggs_ref}
\end{figure*}

Figure \ref{feh_waggs_ref} shows the retrieved values of [Fe/H] compared to the reference from literature. It is noticeable that, despite a small systematic shift ($\sim$ 0.2 dex), the light-weighted and mass-weighted values show a good agreement with the literature. [Fe/H]$_{\rm SSP}$ and the [Fe/H]$_{{\rm max}\ x_j}$, on the other hand, show a bigger scatter, up to 2.0\,dex. 

\begin{figure*}
\centering
	\includegraphics[scale=0.9]{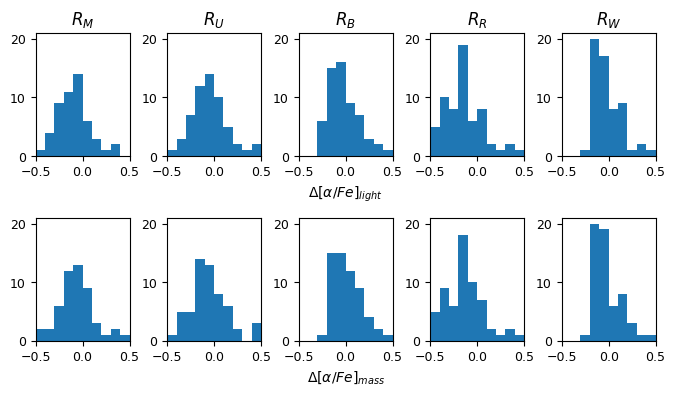}
    \caption{Distributions of $\Delta [\alpha/{\rm Fe}] = [\alpha/{\rm Fe}]_{\rm Starlight} - [Mg/{\rm Fe}]$. Upper row shows the histograms of light-weighted $\Delta [\alpha/{\rm Fe}]$, while the lower row shows the same distributions for mass-weighted values (see Section \ref{different_ages}). Each column represents one of the wavelength ranges tested in this work (see Table \ref{tab_ranges}).}
    \label{delta_alpha}
\end{figure*}

Due to the smaller dynamical range of the \afe\ values, we opted to show the results as the distribution of $\Delta $\afe $=$ \afe$_{\rm Starlight} - $\afe$_{\rm Reference}$. We can see in Figure \ref{delta_alpha} and in Table \ref{delta_parameters} that the retrieved values are closer to the reference values when the fit is performed using the $R_W$ range, not showing significant differences between the light-weighted and the mass-weighted values.

\subsection{\label{sec_errors}Evaluating uncertainties in the retrieved parameters}

The sources of uncertainties are multiple: in the reference values, from the models, from the observations and from the method; and not straightforward to access. 

The uncertainties in the reference values reported in literature are typically small (see Table \ref{ref_table}). 
In order to evaluate the uncertainties in our results due to the quality of the observations (SNR), we used a similar method to the one presented in \citet{cid+05}.
Using the flux uncertainties in the second extensions of the publicly available FITS files for WAGGS library, we created twenty perturbed spectra of each observation, perturbing each flux value ${\rm flux}_j$ with a random value inside its error bar ${\rm err}_j$ (Equation \ref{perturbed}). We then fitted all 20 perturbed observed spectra using the same method and models described above. The uncertainty $\sigma_P$ of each parameter \textit{P} is given by the standard deviation of the values obtained from the 20 realisations (Equation \ref{stddev}). 

\begin{center}
\begin{equation}
    \mbox{flux}_j \rightarrow \mbox{flux}_j + random[-1,1] \times \mbox{err}_j
    \label{perturbed}
\end{equation}

\begin{equation}
    \sigma_P = \sqrt{\frac{\sum_{\substack{i=1}}^{20}{(P_i - \langle P \rangle)^2}}{20}}
    \label{stddev}
\end{equation}
\end{center}

To optimise computing time, we evaluated these uncertainties performing the spectral fits in the interval $R_W$ (4828--5364 \AA), favoured by \citet{walcher+09}. We computed the standard deviation of the parameters retrieved from the 20 realisations, and report their median values in table \ref{tab_uncertainties}. These values can be seen as typical observational uncertainties due to the SNR of the WAGGS sample, while individual uncertainties will vary from spectrum to spectrum. In any case, these simulations show that the typical observational errors are small.

\begin{table}
\caption{Typical standard deviations of the parameters obtained from perturbing the observed spectrum according to its error spectrum.}
\begin{center}
\begin{tabular}{ll}
    \hline
    \textbf{Parameter} & \textbf{median($\sigma$)} \\  
     \hline
    $\log(t_{\rm light})$ & 0.002\\
    $\log(t_{\rm mass})$  & 0.002\\
    \feh$_{\rm light}$    & 0.02 \\
    \feh$_{\rm mass}$     & 0.03 \\
    \afe$_{\rm light}$    & 0.02 \\
    \afe$_{\rm mass}$     & 0.02 \\
     \hline
     \label{tab_uncertainties}
\end{tabular}
\end{center}
\end{table}

Another source of statistical uncertainty not accounted for in the previous estimation is the stochastic sampling of short lived stellar evolutionary phases, due to the finite stellar mass inside the field of view \citep[e.g.][]{daSilva+12}. In a general sense, one should trust more the results for clusters with larger enclosed mass, as given in column 15 of Table 1 from \citet{waggs1}. We investigated if there is a relation between the residuals of the retrieved parameters and the mass enclosed, but no correlation was found.

\subsection{\label{sec_schiavon}Uncertainties coming from different observations of the same clusters}

\begin{figure}
\centering
	\includegraphics[width=0.8\columnwidth]{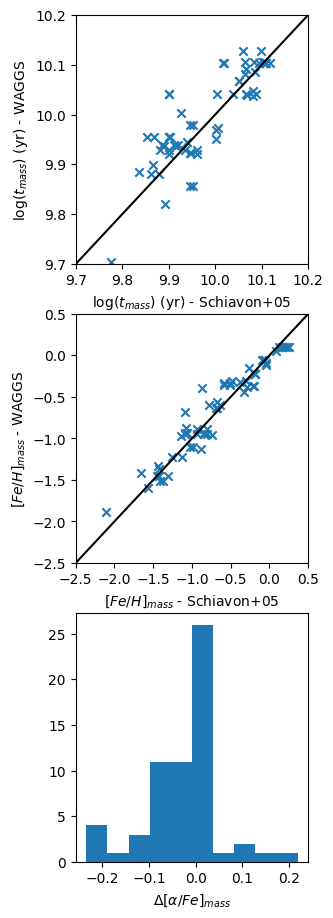}
    \caption{Values for mass-weighted ages and [Fe/H] (upper and middle panels, respectively) retrieved from the WAGGS library versus the values retrieved when fitting spectra from \citet{schiavon+05}'s library. The identity line x=y is shown in solid black to guide the eye. The third panel shows the distribution of $\Delta [\alpha/{\rm Fe}] = [\alpha/{\rm Fe}]_{\rm WAGGS} - [\alpha/{\rm Fe}]_{\rm Schiavon}$.}
    \label{waggs_vs_sch}
\end{figure}

We repeated the spectral fitting in the range $R_W$ using instead the integrated globular spectra from \citet{schiavon+05}, for the 35 globular clusters in common with WAGGS. Figure \ref{waggs_vs_sch} compares the parameters obtained from the two datasets. It is noticeable that the values obtained from the two sets correlate well, with most of the fits showing discrepancy values smaller than 0.1 for log(ages), 0.3 for \feh\ and 0.1 for \afe. 

The source of the deviations can be two-fold: the two data sets been observed with different observing strategies and data reduction pipelines; and
the differences in resolutions, \citet{schiavon+05} spectra having lower resolution (FWHM = 3.1\AA) compared to WAGGS convolved to MILES resolution (FWHM = 2.5\AA).

To verify these possibilities, we convolved WAGGS spectra to FWHM = 3.1\AA\ and run new fits in the $R_W$ range. We evaluate that the differences due to different resolutions are: 
--0.015 $\pm$ 0.057 in  $\log(t_{\rm mass})$, 
--0.014 $\pm$ 0.063 in \feh$_{\rm mass}$, and 
0.015 $\pm$ 0.057 in \afe$_{\rm mass}$. 
In comparison, the differences in parameters between WAGGS (at FWHM = 3.1\AA) and Schiavon data are: 
--0.005 $\pm$ 0.051 in  $\log(t_{\rm mass})$, 
--0.025 $\pm$ 0.167 in \feh$_{\rm mass}$, and 
0.035 $\pm$ 0.078 in \afe$_{\rm mass}$.

We therefore conclude that different observing strategies  and/or data reduction process are comparable to the effect of changing the spectral resolution. Still, the different observations seem to be dominant effect on the dispersion of abundance parameters seen in Fig. \ref{waggs_vs_sch}.  

\subsection{Summary of results}

Table \ref{delta_parameters} shows the median $\pm$ the interquartile range (IQR) of each $\Delta P = P_{\rm Starlight} - P_{\rm Reference}$, where $P$ represents any of the retrieved parameters. 
We see that reddening values are better constrained when more information is used (in our case, the $R_M$ range).
When it comes to ages, the mass-weighted-$R_B$ values are the ones with less deviation from the isochrone (reference) values. 
The closest match to the reference metallicities is reached in the mass-weighted case, using the $R_R$ range in the fit.
For [$\alpha$/Fe], $R_W$ returns the more consistent results, not showing a significant variation between light and mass weighted values. 

The values shown in Table \ref{delta_parameters} can guide the user in the choice of the wavelength range to use for his/her own application. This choice will be dependent on each science case: e.g., if the main goal is to obtain reliable \afe values and a larger uncertainties in ages are manageable, the interval $R_W$ will be favoured; conversely, if maximising the accuracy of $\Delta\log(t_{\rm mass}$ is the goal, we suggest $R_B$. Unfortunately we cannot suggest a single wavelength range for all cases.

The ranges suggested here are admittedly based on the WAGGS spectra, but for a more general application one could inspect Table \ref{features} to verify which spectral features are driving the results. By looking at Table \ref{features} alone one would expect $R_U$ to be the range most sensitive to age, but we interpret that the lower SNR in this region causes the best age-determination range to shift to $R_B$. The range $R_W$ is a narrow one around the Mgb triplet, so it is not surprising that it performed best for deriving \afe\ values. Regarding metallicity, the range which performed best on average is $R_R$, basically devoid of age features (other than H$\alpha$), even though with a relatively large IQR.

\begin{table*}
\caption{Median values of $\Delta$parameters for each given wavelength range. $\Delta P$ = $P_{\rm Starlight} - P_{\rm Reference}$ where $P$ represents each parameter. Columns 2 -- 6 indicate the fitted wavelength range, as defined in table \ref{tab_ranges}. }
\begin{tabular}{l|c|c|c|c|c}
    \hline
     Parameter & \textbf{$R_M$} & \textbf{$R_U$} & \textbf{$R_B$} & \textbf{$R_R$} & \textbf{$R_W$} \\
     \hline
     {$\Delta\log(t_{\rm light})$} & -0.44 $\pm$ 0.45 & -0.27 $\pm$ 0.23 & -0.40 $\pm$ 0.41 & -0.56 $\pm$ 0.41 & -0.19 $\pm$ 0.29 \\ 
     {$\Delta\log(t_{\rm mass})$} & -0.19 $\pm$ 0.34 & -0.19 $\pm$ 0.16 & -0.06 $\pm$ 0.17 & -0.17 $\pm$ 0.18 & -0.14 $\pm$ 0.22 \\ 
     {$\Delta{\rm [Fe/H]}_{\rm light}$} & 0.33 $\pm$ 0.28 & 0.34 $\pm$ 0.21 & 0.24 $\pm$ 0.20 & 0.25 $\pm$ 0.27 & 0.18 $\pm$ 0.24 \\ 
     {$\Delta{\rm [Fe/H]}_{\rm mass}$} & 0.23 $\pm$ 0.32 & 0.39 $\pm$ 0.29 & 0.34 $\pm$ 0.26 & 0.09 $\pm$ 0.39 & 0.22 $\pm$ 0.27 \\
     {$\Delta{\rm [\alpha/Fe]}_{\rm light}$} & -0.10 $\pm$ 0.19 & -0.05 $\pm$ 0.22 & -0.05 $\pm$ 0.23 & -0.17 $\pm$ 0.22 & -0.05 $\pm$ 0.17 \\
     {$\Delta{\rm [\alpha/Fe]}_{\rm mass}$} & -0.08 $\pm$ 0.20 & -0.06 $\pm$ 0.22 & -0.02 $\pm$ 0.22 & -0.14 $\pm$ 0.23 & -0.05 $\pm$ 0.17 \\
     {$\Delta A_V$} & 0.10 $\pm$ 0.30 & -0.25 $\pm$ 0.50 & 0.12 $\pm$ 0.28 & 0.31 $\pm$ 0.40 & 0.14 $\pm$ 0.38 \\
     \hline
     \label{delta_parameters}
\end{tabular}
\end{table*}

\section{Discussions}
\label{sec_discussions}

\begin{figure}
\centering
	\includegraphics[width=\columnwidth]{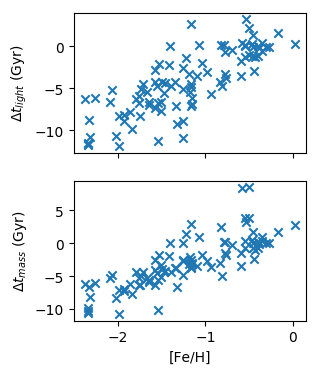}
    \caption{$\Delta t = t_{\rm Starlight} - t_{\rm iso}$ values (light-weighted in upper panel, mass weighted in lower panel) versus [Fe/H], fitting the $R_W$ interval. It is noticeable that there is a clear correlation between the differences between fitted and isochrone ages with the metallicity.}
    \label{deltat_feh}
\end{figure}

\paragraph*{On blue HB stars and blue stragglers}
Arguably, the most striking failure in retrieving the parameters from our exercise in spectral fitting are the ages of the low metallicity systems (figure \ref{ages_linear}). We better highlight the dependence of the problem with metallicity in Figure \ref{deltat_feh}, which shows 
$\Delta t_i = t_i - t_{iso}$ (being $t_i$ the different ages defined in section \ref{different_ages}) vs. \feh, 
for the range $R_W$. The classical interpretation of blue-light excess in Galactic old populations is the presence of extended HB morphologies or blue-stragglers, when unaccounted for in SSP models. The "excess" of blue light coming from these stars could be interpreted by the algorithm as young bursts of star formation, lowering the inferred ages of the integrated spectra. 
These effects have been discussed in a myriad of papers \citep[e.g.][]{pacheco_barbuy95, lee+00, maraston_thomas00, peterson+03, schiavon+04b}. 

In particular in the context of spectral fitting, we refer the reader to the work of \citet{koleva+08} and \citet{ocvirk10}.  \citet{koleva+08} were able to reproduce CMD ages for 35 out of 40 clusters, using hot stars together with the SSPs when needed, to mimic the effect of the HB morphology (or blue stragglers). \citet{ocvirk10} proposes that any young starburst superimposed on an old stellar population in this range \feh = $[-2, -1.2]$ could be regarded as a modelling artefact, if it weighs less than 12\% of the optical light. The work by \citet{conroy+18} illustrates a similar result (see their Figure 15), in a sample with \feh\ between $[-1.5, +0.3]$. It remains to be investigated in a future work if this effect can explain all the deviant cases found in this work.

\paragraph*{Quality of the SSP model at low metallicities}

The SSP models used in this work \citep{vazdekis+15} are built upon the empirical stellar library MILES \citep{MILES1}. The coverage of the Hertzsprung–Russell diagram (HRD hereafter) in empirical stellar libraries in general is known to be poorer towards lower metallicities, due to observational constraints (see e.g. \citealt{coelho09a_proc} and Section 3.2 in \citealt{vazdekis+10}).

On the one hand, according to the analyses in \citet{vazdekis+10}, old models ($>\sim10$\,Gyr) can be safely used also at low metallicity. In this aspect the poorer coverage at low metallicites should not be a reason for the discrepant ages observed here. On the other hand, the recent work by \citet{coelho+19} investigated the effect that the HRD coverage has on the SSP model predictions. In their figure 15 and table 7, there is evidence that the sparse coverage of HRD imply in an underestimation of the inferred ages. The authors observe a median effect of $\Delta \log(t)\sim 0.11$, observed in systems with metallicity between \feh\ $-1$ and $0$. It remains an open possibility, then, if this effect could partly account for the age discrepancies observed in this work. 

\paragraph*{The choice of isochrones}

Ideally, one would want the SSP models to adopt the same isochrones as the ones used to derive the reference ages. Or the other way around, obtain from literature age estimations using the same isochrones as the ones used in the SSP models.  This is not the case here, since the SSP models are using isochrones from \citet{pietrinferni+04,pietrinferni+06}, while the reference ages come
from mixed sources (see table \ref{ref_table}). 

The review from \citet{gallart+05} compares several stellar evolutionary libraries from literature, and quantify how they (dis)agree when used to infer age, metallicity, or distance of a population.  
Ages predicted by different models agree within $\pm$0.01 \,Gyr at young ages ($\sim$0.1 Gyr), and differ by up to 1\,Gyr in the intermediate and old age regimes. We therefore assume that different isochrones can account for a spread of 1\,Gyr in inferred ages, which may be important to intermediate-age populations.

On the other hand, one may argue that using the same isochrones does not necessarily help: the isochrones used to measure the age from CMDs could make a good prediction of the main-sequence turn-off (MSTO) but not of later stellar evolution phases (such as the horizontal branch or red giant branch), which may dominate over the MSTO in the integrated light.



\paragraph*{Elusive \afe}

The \afe\ is an important chemical marker in stellar populations, working as a cosmic clock for the history of star formation \citep[e.g.][]{matteucci03}. It has been measured in galaxies for more than a decade using spectral indices \citep[e.g.][]{thomas+05}, but our results seem to indicate that it can be elusive to measure from spectral fitting. Qualitatively similar results are shown in Figure 15 in \citet{conroy+18}, where measuring individual abundances other then \feh\ show considerable scatter. We could partly attribute the difficulties to the dynamical range -- considerably smaller than the other parameters -- from $\sim -0.2$  to $\sim +0.5$ in the observations, and only two values available in the models (0.0 and +0.4). In any case, the results in Table \ref{delta_parameters} and Figure \ref{delta_alpha} indicate that uncertainties similar or larger than 0.2\,dex are to be expected, and that fitting a narrow wavelength range is more indicated (at the expense of worsening the other results).

\paragraph*{Example of an extreme case and uncertainties:}

We discussed possible uncertainties coming from the observations in Sections \ref{sec_errors} and \ref{sec_schiavon}. Estimating errors in the models and method are considerably more difficult. In the case of the SSP fits, we can follow a procedure similar to the one adopted in \citet[][see section 4.2]{charlot+02}, namely, considering that all models within $\Delta \chi^2 < 1$ around the best $\chi^2$ could also be picked as a good fit. 

In our analysis this approach is limited to the SSP fits (because \textsc{Starlight} outputs the $\chi^2$ of all SSP fits, but not of all multi-population fits of the Markov-Chain), and should be regarded with care given the biases that we are aware to exist in the SNR estimation (see section \ref{obsdata}). 

In any case, as a test-case, we choose to apply this scheme to the the results from NGC6397, which falls into the worst-case scenario of our analysis of ages: isochrone fitting returns an age of 13.5 Gyr, while the retrieved light-weighted, mass-weighted and SSP-equivalent age values are about 1.0, 5.0 and 2.0 Gyr, respectively.

If one estimates the age uncertainty from the criteria $\Delta \chi^2 < 1$, the acceptable ages for NGC6397 
range from 1.25 to 7.50\,Gyr for SSPs with 
\feh\ $=  -2.27$,
 and from 1.25 to 4.50\,Gyr for SSPs with \feh\ $=  -1.79$ (bracketing the literature value of \feh $= -2.0$ for this cluster).
Although not large enough to bring the result in agreement with the isochrones age of the cluster, the intervals are rather large, suggesting that indeed the models and/or method may not be able to clearly distinguish the age parameter. For cases less extreme than NGC6397, this could bring the results in agreement with literature.

\section{Conclusions} \label{sec_conclusion}

In this work we obtained reddening $A_V$, ages, \feh\ and \afe\ from spectral fitting of integrated spectra of globular clusters \citep{waggs1}, simulating a procedure which is routinely applied to the analysis of galaxy spectra. We repeated the analyses for different wavelength ranges, and compared the retrieved parameters with reference values compiled from literature (obtained from CMD analysis and high-resolution spectroscopy), to decide which ranges are more robust. Our main conclusions are summarised below. 

The intervals which best reproduced fiducial values of reddening, ages, \feh\ and \afe\ are different (respectively $R_W$ 3540--7409, $R_B$ 4170--5540, $R_R$ 5280--7020 and $R_W$ 4828-5364). 
For all parameters, changing the wavelength range of the spectral fit changes the parameters inferred. Reddening is the only parameter which is favoured by the use of the widest possible interval. On the other limit, \afe\ is better retrieved when the narrowest range is fitted.

Ages are poorly constrained for metal-poor objects, in accordance with previous results in literature. The correlation between age residuals (retrieved -- literature) with metallicity seems to be consistent with the effect of blue HB stars or blue stragglers. Alternatively, the sparser coverage of the HRD by the models at low metallicities may also play a role.  

The explanations of our findings can be multi-fold, and one can argue that it may be due to the choice of models or code employed. But we stress that our results are qualitatively similar to what other work found with different models and different fitting codes \citep[e.g.][which employed codes that do not use the continuum information]{walcher+09, cezario+13}. Admittedly, the result that different wavelength ranges will return different parameters is an uncomfortable one. Moreover, our results do not conform with the common assumption existent in the spectral fitting community where "the more pixels the better". That drives us back to the decades-old work on defining spectral indices, where the features driving the stellar population parameters were determined almost in an artisanal manner. We make the hypothesis that to use more pixels can bring more \emph{signal} but do not necessarily bring more \emph{information}, in the sense that we want to maximise something similar to the SNR concept (increase the signal without increasing the noise). Using the case of \afe\ as an example, centring the fit around Mg$_b$ triplet feature may maximises the information, which otherwise would be diluted in a larger wavelength range.

In essence we conclude that a user of spectral fitting codes interpreting galaxies will not obtain the best result necessarily by using more data (wider wavelength ranges). 

    

\section{DATA AVAILABILITY}

 Value-added catalogues produced in this work using spectral fitting are available in the article's online supplementary material. Instructions on how to acquire the public data herein used are described in Sec. \ref{obsdata}.

\section*{Acknowledgements}
GG acknowledges support from Coordenação de Aperfeiçoamento de Pessoal de Nível Superior (CAPES).
PC acknowledges support from Conselho Nacional de Desenvolvimento Cient\'ifico e Tecnol\'ogico
(CNPq 310041/2018-0) and Funda\c c\~ao de Amparo \`a Pesquisa do Estado de S\~ao Paulo
(project 2018/05392-8).
GG and PC thank Ariel Werle, Roberto Cid Fernandes, Jacopo Chevallard, Gustavo Bruzual and Stephane Charlot for the several discussions that enriched the development of this work.




\bibliographystyle{mnras}
\bibliography{main} 


\bsp	

\label{lastpage}
\end{document}